\documentclass[useAMS,usenatbib,epsfig,epsf]{mnras}
\usepackage{times,epsfig}
\usepackage{natbib}
\usepackage[T1]{fontenc}
\usepackage{ae,aecompl}
\title[K2 observations of PG\,1142-037]
{A pulsation analysis of \emph{K2} observations of the subdwarf B
star PG\,1142-037 during Campaign 1: A subsynchronously rotating ellipsoidal
variable }

\author[M.D. Reed et al.]{
 M.\,D.\,Reed$^1$\thanks{E-mail:MikeReed@missouristate.edu},
A.\,S.\,Baran,$^2$, R.\,H.\,\O stensen$^3$, J.\,H.\,Telting$^4$, J.\,W.\,Kern$^1$, 
S.\,Bloemen$^5$ \newauthor
P.\,Blay$^{6,4}$, T.\,Pursimo$^4$, T.\,Kuutma$^{4,7}$, 
 D.\,Slumstrup$^{4,8}$, M.\,Saajasto$^{4,9}$, L.\,D.\,Nielsen$^4$, 
\newauthor J.\,Harmanen$^{4,10}$, 
A.\,J.\,Winans$^1$, H.\,M.\,Foster$^1$,
and L.\,Rowe$^1$\\
 $^1$Department of Physics, Astronomy and Materials Science,
 Missouri State University, 901 S. National, Springfield, MO 65897, USA \\
$^2$Suhora Observatory and Krakow Pedagogical University, ul. Podchor\c{a}\.{z}y
ch 2,30-084 Krak\'{o}w, Poland\\
$^3$Instituut voor Sterrenkunde, KU Leuven, Celestijnenlaan 200 D, 3001 Leuven, 
Belgium\\
$^4$Nordic Optical Telescope, Rambla Jos\'e Ana Fern\'andez P\'erez 7, 38711 
Bre\~na Baja, Spain\\
$^5$ Department of Astrophysics/IMAPP, Radboud University Nijmegen, PO Box 9010, NL-6500 GL Nijmegen, the Netherlands\\
$^6$ Instituto Astrofisico de Canarias, Spain\\
$^7$ Tartu Observatory, Observatooriumi 1, 61602 T\~oravere, Estonia\\
$^8$ Stellar Astrophysics Centre, Department of Physics and Astronomy, Aarhus University,
Ny Munkegade 120, DK-8000 Aarhus C, Denmark\\
$^{9}$ Department of Physics, P.O. Box 64, FI-00014, University of Helsinki, Finland\\
$^{10}$ Tuorla Observatory, Department of Physics and Astronomy, University of Turku,
V\"ais\"al\"antie 20, 21500 Piikki\"o, Finland}

\date{Accepted
      Received }

\begin{document}

\maketitle

\begin{abstract}
We report a new subdwarf B (sdB) pulsator, PG\,1142-037, discovered
during the first full-length campaign of \emph{K2}, the two-gyro mission of
the \emph{Kepler} space telescope. Fourteen periodicities have been detected
between 0.9 and 2.5\,hours with amplitudes below 0.35\,ppt.
We have been able to associate all of the pulsations with low-degree,
$\ell\,\leq\,2$ modes. Follow-up spectroscopy of PG\,1142 has revealed
it to be in a binary with a period of 0.54\,days. Phase-folding
the \emph{K2} photometry reveals a two-component variation including both Doppler
boosting and ellipsoidal deformation.
Perhaps the most surprising and interesting
result is the detection of an ellipsoidal, tidally distorted variable
with no indication of rotationally-induced pulsation multiplets. This
indicates that the rotation period is longer than 45 days, even though
the binary period is near 13 hours.
\end{abstract}


\section{Introduction}
\emph{K2} is the follow-up to the very successful \emph{Kepler} space telescope 
mission, using the two surviving reaction wheels to stabilise the pointing
\citep{howell14a}. In this configuration, the spacecraft can reliably track 
targets at coordinates falling close to the ecliptic plane, where solar 
radiation pressure and regular thruster firings help keep the pointing to 
within a few pixels.
Fields are mostly determined by pointing stability demands, in
coordination with
observer-proposed programmes. An observing \emph{campaign} on any given 
field can last up to
around 90\,days. As with the original \emph{Kepler} 
mission, observers must propose individual targets within the selected fields 
for long cadence (LC, 30\,minutes) or short cadence (SC, 1\,minute) observations.

Our interest is asteroseismology of pulsating subdwarf B stars (sdB; also
known as extreme horizontal branch stars), for which \emph{Kepler} data
have proven to be extremely useful. The near-continuous monitoring afforded
by \emph{Kepler} has revolutionized our ability to associate pulsation modes with
pulsations. Subdwarf B stars pulsate in both gravity (V1093\,Hya) and
pressure (V361\,Her) modes with periods near an hour or a few minutes,
respectively. Most of the \emph{Kepler}-observed sdB pulsators (sdBV)
were $g-$mode pulsators \citep{roy10b,roy11b}, some also with
$p-$mode periodicities. The detection of evenly-spaced
$g-$mode periods \citep{reed11c} and rotationally-induced frequency
multiplets \citep{baran12a} has allowed most periodicities
to be associated with modes, predominantly of low-degree $\ell\,\leq\,2$.
Discoveries from these data include slow rotation periods 
\citep[see Table 2 of][]{reed14} even in $<1$\,day binaries 
\citep{pablo11,pablo12,baran12c}, including one with differential
rotation \citep{mdr15}; some high-degree ($\ell\,\geq\,3$)
modes, including $\ell\,\geq\,8$ \citep[See Fig.\,9 of][]{jht14};
some oscillations with stochastic properties \citep{roy14b}; 
and while evenly-spaced
periods indicate smoother core-envelope transitions than anticipated
\citep[see discussion in][]{reed14}, at least two sdB stars have trapped modes
\citep{mdr15,roy14a}.

In preparation for the K2 mission, we made a selection of UV-bright targets 
falling near the ecliptic plane based on photometry from the GALEX satellite 
\citep{GALEX}. The brighter of these targets were observed spectroscopically from
the INT, KPNO, NOT, and NTT 
starting in January 2014. The list contains 715 targets, 
and observations are ongoing, focusing on proposed K2 campaign fields. 

Here we report the discovery of a variable star from our analysis of K2's
first campaign (C1) photometry, which spanned 93 days during May -- August of 2014.
The star, EPIC 201206621, is identified with PG 1142-037
(hereafter PG\,1142)  from the Palomar-Green survey which classified it
as sdB-O with B$\,=\,15.67$ \citep{PGSurvey}. It was observed using
RC-Spec at the Kitt Peak National Observatory 4\,m telescope during February 
of 2014 as part of our spectroscopic program described above.
Two 900\,second spectra were obtained which were sufficient to determine
$T_{\rm eff}\,\sim\,27000\,K$
and $\log g\,\sim\,5.30$ using the same LTE models as described in \S 2.1, 
which placed it within the $g-$mode instability region. 
During 2014, PG\,1142 was observed for 2.3\,hours to search for photometric 
variability by MDR. No pulsations were detected to a noise level of 
8.1\,parts-per-thousand (ppt). 
PG\,1142 was listed in
\citet{reed04} as not having a main sequence companion earlier than M2
using 2MASS data.

\section{Observations and data processing}
PG\,1142 was observed by \emph{K2} 
in SC mode which summed nine images
into $58.8\,s$ integrations. We downloaded pixel array files
($15\times 14$) from the 
Mikulski Archive for Space Telescopes. Unlike the main \emph{Kepler} mission,
K2 does not yet supply processed lightcurves, so 
investigators must extract fluxes from pixel data.

We used a set of custom scripts for extracting the
light curve from the pixel files. First, we employed the {\sc PyFits} 
library within {\sc Python}
to read fits tables and pull out timestamps and fluxes. We limited data points
to those having quality flag values of zero, 
therefore we avoided points affected by the
onboard systematics. Second, we extracted light curves and stellar profile positions
using the standard {\sc iraf} aperture photometry package {\sc phot} 
after converting the pixel files into
individual images using the {\sc KEPIMAGE} program. This had the advantage
that we could choose a small aperture which would follow
the image center using the {\sc IMCENTROID} task.
Since no flat fielding is applied to the data, we de-correlated the fluxes with 
stellar profile positions to
remove the most disturbing feature present in K2 data, the roughly
six hour periodicity caused
by thruster firings to keep the spacecraft accurately pointed.This de-correlation 
was done on 10-day chunks of data.
Finally, we removed long-term ($>6$\,days) trends with Akima splines, 
normalized using the original median flux in each 10-day chunk,
subtracted 1, and multiplied by 1000 to have differential fluxes in 
part-per-thousand (ppt).

We also processed the data using the {\sc pyke} \citep{still12} software packages
{\sc kepmask}, {\sc kepextract}, {\sc kepflatten}, and
{\sc kepconvert} as well as the
\emph{K2}-developed {\sc kepsff}. While these routines work well, we 
found that our techniques described above 
provide lower noise and so we did not use {\sc pyke}-processed data.

\subsection{Spectroscopic Parameters}
Once pulsations were detected, we started observing PG\,1142
spectroscopically as part of our campaign dedicated to
investigating the binary status of pulsating hot subdwarfs observed by
Kepler \citep[e.g.][]{jhtsdob6}. In February, March, and June 2015 we
obtained a total of 24 radial-velocity (RV) measurements of PG\,1142, as
listed in Table\,\ref{tab02}.

\begin{table}
\caption{Spectroscopic observations of PG\,1142.}
\label{tab02}
\begin{tabular}{cccccc} \hline
Date & BJD & S/N & RV & Error \\ \hline
2015-02-02T03:36:01.8 & 2457055.6539860 & 52.3 & -20.037 & 5.674 \\
2015-02-02T06:05:15.2 & 2457055.7576212 & 66.3 & -92.296 & 7.049 \\
2015-02-04T04:32:01.6 & 2457057.6930173 & 48.9 & 66.725 & 4.207 \\
2015-02-04T06:23:54.7 & 2457057.7707205 & 55.7 & 21.640 & 7.752 \\
2015-02-06T01:04:41.9 & 2457059.5491660 & 44.6 & -97.101 & 6.922 \\
2015-02-07T01:16:57.7 & 2457060.5577485 & 27.4 & -66.534 & 10.827 \\
2015-02-14T06:38:28.3 & 2457067.7814588 & 54.3 & -47.224 & 5.931 \\
2015-02-15T01:20:05.8 & 2457068.5604078 & 44.8 & 51.260 & 6.892 \\
2015-02-15T05:12:26.7 & 2457068.7217691 & 27.1 & -68.086 & 10.040 \\
2015-03-01T23:31:21.3 & 2457083.4855331 & 46.9 & -33.843 & 11.161 \\
2015-03-02T02:33:37.4 & 2457083.6121128 & 50.0 & 74.737 & 7.231 \\
2015-03-04T01:51:41.2 & 2457085.5830472 & 43.6 & -91.444 & 5.336 \\
2015-03-05T00:44:01.7 & 2457086.5360873 & 40.5 & -64.897 & 6.872 \\
2015-03-12T04:46:23.1 & 2457093.7045330 & 51.8 & -82.345 & 5.677 \\
2015-03-16T00:26:21.5 & 2457097.5239989 & 55.0 & -59.301 & 4.197 \\
2015-03-17T02:40:42.8 & 2457098.6173076 & 25.7 & -49.665 & 7.195 \\ \hline
2015-06-12T21:28:07.3 & 2457186.3951956 & 46.4 & 73.508 & 7.843 \\
2015-06-12T22:46:09.4 & 2457186.4493808 & 55.7 & 80.049 & 6.347 \\
2015-06-16T21:05:47.5 & 2457190.3793016 & 37.0 & -29.492 & 6.194 \\
2015-06-16T21:25:36.2 & 2457190.3930589 & 35.3 & -25.492 & 8.038 \\
2015-06-16T22:28:48.9 & 2457190.4369509 & 37.7 & -55.445 & 6.903 \\
2015-06-16T22:49:07.5 & 2457190.4510538 & 43.5 & -62.600 & 6.493 \\
2015-06-17T21:16:19.9 & 2457191.3865230 & 26.3 & 52.608 & 11.179 \\
2015-06-17T23:14:12.0 & 2457191.4683685 & 56.4 & -21.937 & 6.596 \\ \hline
\end{tabular}
\end{table}

All low-resolution spectra were collected with the ALFOSC spectrograph at
the Nordic Optical Telescope (NOT), using a 0.5 arcsec slit with the new
high-efficiency volume-phased holographic grism \#18, covering the approximate wavelength range
of 3530-5200\,\AA, with resolution R$\sim$2000 (or 2.2\AA) and dispersion of
0.82\,\AA/pix.
The exposure time used was 900 seconds, except for one spectrum that was
exposed 1500 seconds. The S/N of the spectra ranges from
26 to 66, with a median value of 46, depending on observing conditions.

All spectra were processed and extracted using standard {\sc iraf} tasks. Radial
velocities (RVs) were computed with {\sc fxcor}, 
by cross-correlating with a synthetic
template derived from a fit to a mean spectrum of the target (see below),
while using the H$\beta$, H$\gamma$, H$\delta$, H8, and H9 
lines from the observed spectra. After the first iteration, there
was $\sim 4\%$ velocity residuals in phase with the signal and so 
we did an
extra {\sc fxcor} iteration on spectra that were shifted to correct for the
first RV results. This procedure led to a final RV amplitude about 4\%
larger than obtained from the first try. The errors returned by the second
{\sc fxcor} run are more consistent with the RMS from the orbit fit 
(see \S 3.2),
so we used these errors for the analysis in this paper. The median
RV error is $7$\,km\,s$^{-1}$.

The final RVs were adjusted for the position of the target on the slit,
judged from slit images taken just before and after the spectra. Table\,1
lists the observations with their mid-exposure dates, and RV measurements
with the final {\sc fxcor} error.

\begin{figure*}
\centerline{\psfig{figure=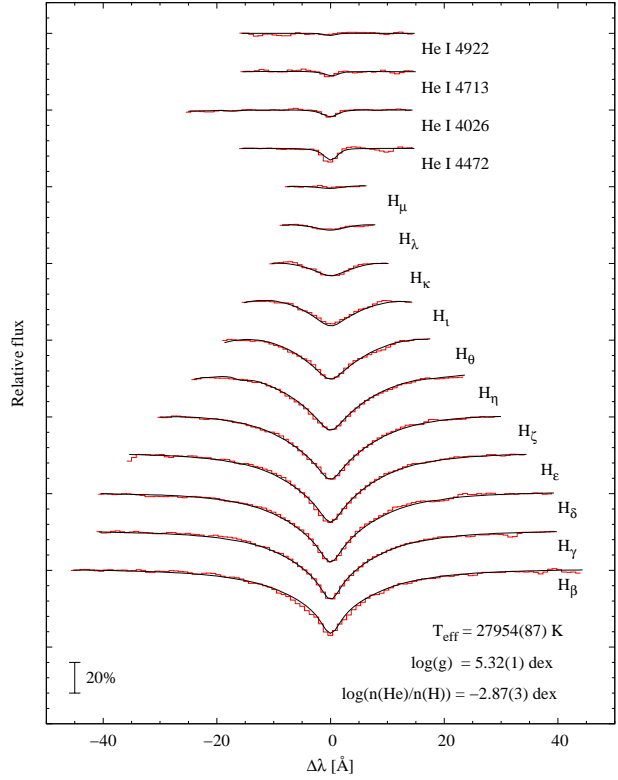,width=\columnwidth} }
\caption{Fit to mean spectrum of PG\,1142. Black lines show the fit while
red ones the mean spectrum. Each line is labelled.}
\label{figspec}
\end{figure*}

We used the mean of the best 12 spectra (each with S/N>45), after shifting each
to remove the orbit, and with final S/N$\sim$180, to obtain the first high
signal-to-noise determination of the atmospheric parameters of PG\,1142.
We determined $T_{\rm eff}$ and $\log g$ from the mean spectrum using the 
H/He LTE
grid of \citet{heber99b} for consistency with \citet{roy10b}.
We used all the Balmer lines from H$\beta$ to H14
and the four strongest He\,I lines for the fit.  We find
$T_{\rm eff}\,=27954\pm 87,K$,
$\log g\,=\,5.32\pm0.01$, and
$\log$(N(He)/N(H))$\,=\,-2.87\pm0.03$.
The errors listed on the measurements are the formal errors of the fit,
which reflect the S/N of the mean spectrum. These values and errors are
relative to the LTE model grid and do not reflect any systematic effects
caused by the assumptions underlying those models. The model fit is shown
in Fig.\,\ref{figspec}.

\section{Data Analysis}
\subsection{Pulsation Analysis}
93\,days of near-continuous observations yields a 1/T resolution of $0.12\,\mu$Hz, which
is $\sim71,000$ independent frequencies up to the Nyquist.
Initially, we used a detection threshold of $4.35\sigma$ (0.147\,ppt) 
as indicated in \citet{bev}
and then $5.2\sigma$ (0.175\,ppt) 
as indicated in \citet{andy15}. However, having processed the
data several times using multiple methods, we noted periodicities which were 
consistently distinguished from random noise. Using the higher detection limit to
construct period sequences (described below), we noticed that the lower amplitude
periodicities also fit into the sequences, and so retained them, even though they
fall below a justifiable detection limit. We include the S/N values in Table\ref{tab01}
and note those frequencies with low values.

\begin{figure*}
\centerline{\psfig{figure=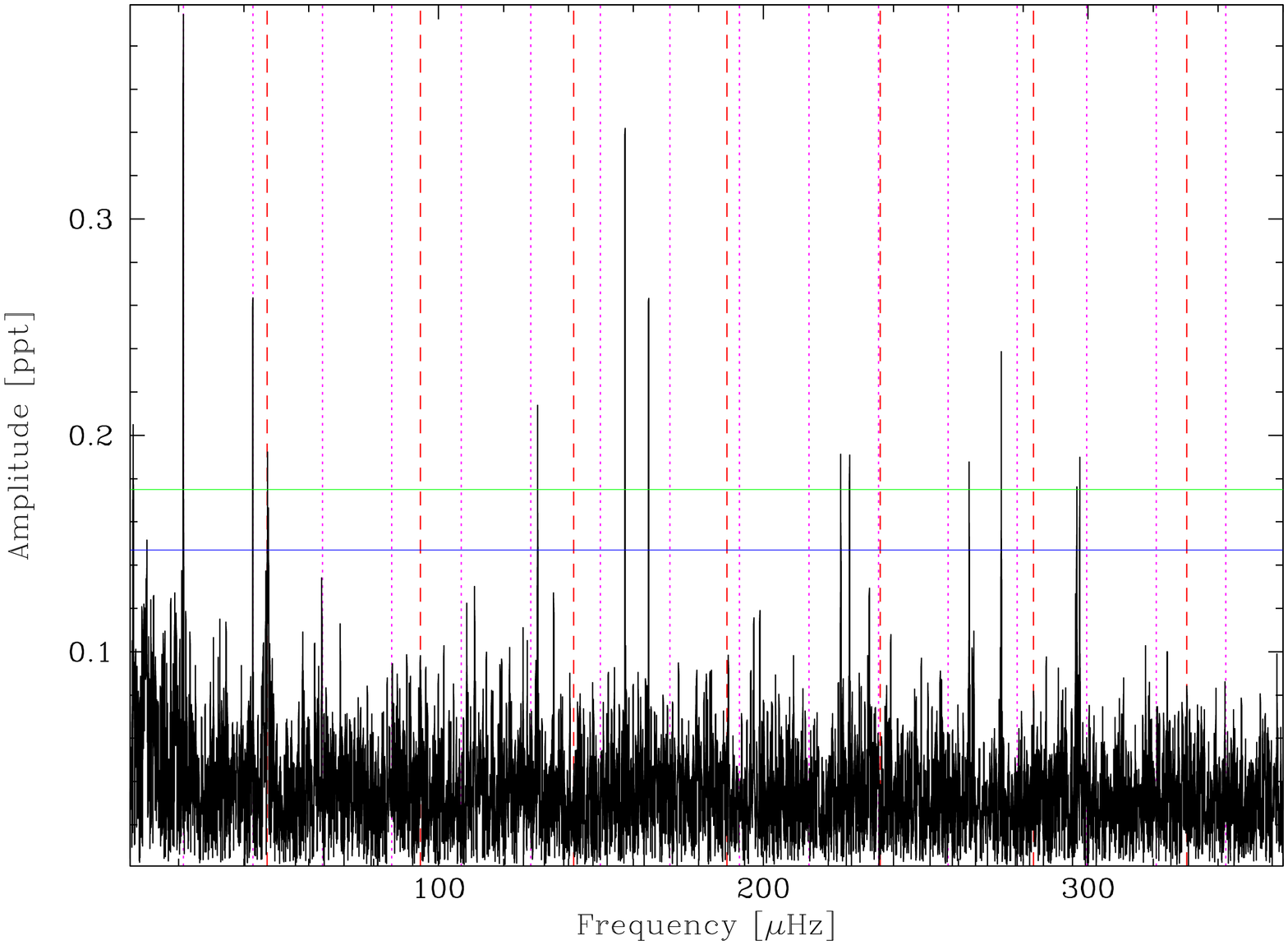,angle=-90,width=\textwidth} }
\caption{Fourier transform of PG\,1142.
Vertical dashed (red) lines indicate the
6\,hr thruster-firing periodicity and overtones and the dotted (magenta) lines
indicate the binary period and its overtones. The horizontal
lines indicate the 4.35 and 5.2$\sigma$ detection limits. }
\label{fig01}
\end{figure*}

\begin{figure*}
\centerline{\psfig{figure=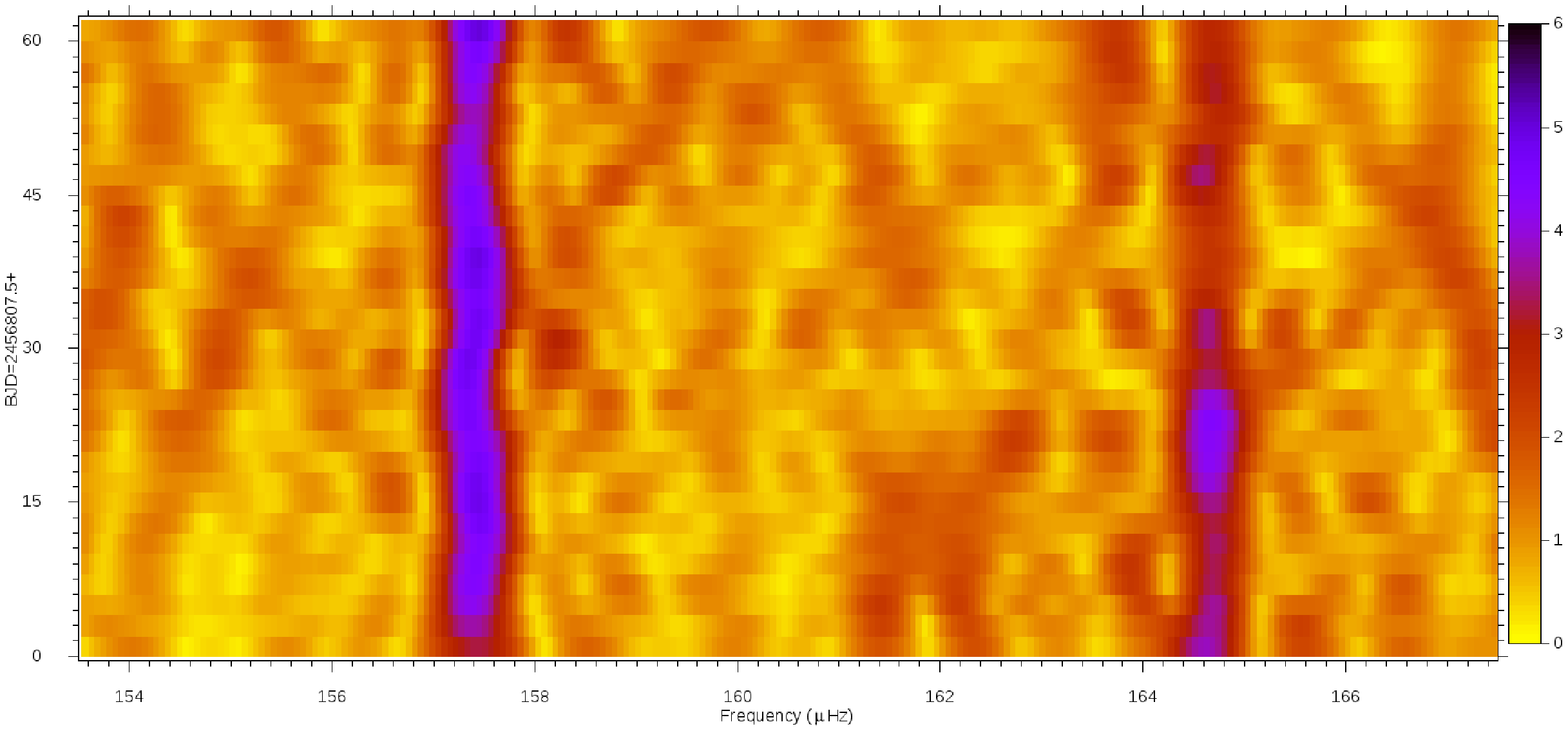,angle=-90,width=\textwidth} }
\caption{Sliding Fourier transform of the two highest amplitude periodicities
showing amplitude variability.
Frequency is on the abscissa and time on the ordinate with color indicating the
amplitude, in $\sigma$ with a scale bar on the right. Each FT spans 20\,days of data
with each element sliding by 2\,days.}
\label{figsft}
\end{figure*}

To determine the pulsation frequencies, amplitudes, and errors,
we fitted Lorentzian profiles to the peaks. This has become our standard procedure when
pulsations have substantial amplitude and/or frequency variability. The
previous method of non-linear least-squares fitting and prewhitening is no
longer effective as it assumes constant amplitudes, phases, and frequencies 
while the observations obviously do not posses these properties. Figure\,\ref{figsft}
shows a sliding FT of the two highest amplitude periodicities, for reference.
We do not wish to imply
that the pulsations are stochastically excited
\citep[see][for discussions]{roy14b,reed07b},
but use this tool as the most effective for determining frequency centers
with the Lorentzian widths providing an indicator of line variability, regardless
of its cause. The fitted periodicities are listed in Table\,\ref{tab01} with other
asteroseismically interesting quantities (described below).

The FT has 12 peaks above a $5.2\,\sigma$ detection limit.
Three of these have long periods at $21.4\,\mu$Hz (12.99\,hrs), its overtone at
$42.8\,\mu$Hz and another at $47.8\,\mu$Hz, which is the residuals of the
thruster firings (not included in Table\,\ref{tab01}). 
Five other frequencies persistently appeared near $4\sigma$ through our various
reductions. All of these occur near period spacing sequences (see discussion below)
and so we include them in Table\,\ref{tab01}.
Figure\,\ref{fig01} shows the FT with noise limits, artefacts, and binary
signals and overtones indicated.

\begin{table*}
\caption{Periods detected for PG\,1142. Column 1 provides an ID, columns 2 and 3 provide
frequencies and periods with errors in parentheses. Column 4 lists the 
amplitude and Column 5 lists
the corresponding signal-to-noise (S/N). Column 6 lists the mode degree with 
columns 7 and 8 listing a relative radial indices.
Columns 9 and 10 list the deviation from the asymptotic sequence. 
No peaks attributed to spacecraft artefacts are listed in this table.
Note: 
$\dagger$ these periodicities have low S/N in our final processing, but match 
asymptotic spacing and so were retained in our table.}
\label{tab01}
\begin{tabular}{lcccccccccc} \hline
ID & Freq & Period & Amp & S/N & $\ell$ & $n_{\ell=1}$ & $n_{\ell=2}$ & $\delta P/\Delta \Pi_1$ & $\delta P/\Delta \Pi_2$ \\
  & ($\mu$Hz) & (sec) & (ppt) & \\ \hline
fA & 21.37 (9) & 46761 (188)	& 0.40			\\
fB & 42.78 (7) & 23377 (39) & 0.26			\\
f1$\dagger$ & 108.67 (9) & 9199.6 (6.9) & 0.12 & 3.5 & 2 &    & 57  & & -0.01  \\
f2$\dagger$ & 111.05 (9) & 9004.7 (6.9) & 0.13 & 3.8 & 1 & 31 &  & -0.037 &   \\
f3 & 130.50 (7) & 7663.1 (4.2) & 0.21 & 6.2 & 1 or 2 & 26 & 47 & -0.047 & 0.006 \\
f4$\dagger$ & 135.41 (9) & 7396.4 (5.0) & 0.13 & 3.8 & 1 & 25 &    &  &-0.04 \\
f5 & 157.38 (9) & 6354.0 (3.7) & 0.34 & 10.0 & 1 & 21 &  & 0.064 &  \\
f6 & 164.65 (8) & 6073.6 (3.0) & 0.26 & 7.6 & 1 & 20 &  & 0.017 &   \\
f7 & 223.82 (9) & 4467.9 (1.8) & 0.19 & 5.6 & 1 & 14 &  & 0.020 &   \\
f8 & 226.55 (8) & 4414.0 (1.6) & 0.19 & 5.6 & 2 &  & 26 &   & -0.103 \\
f9$\dagger$ & 232.63 (12) & 4298.7 (2.2) & 0.13 & 3.8 & 2 &  & 25 &   & 0.148 \\
f10 & 263.36 (8) & 3797.1 (1.2) & 0.19 & 5.6 & 2 &  & 22 &   & -0.111 \\
f11 & 273.25 (8) & 3659.7 (1.0) & 0.24 & 7.1 & 1 or 2 & 11 & 21 & 0.002 & -0.003 \\
f12$\dagger$ & 296.12 (9) & 3377.2 (1.0) & 0.13 & 3.8 & 1 & 10 &  & -0.050 &  \\
f13 & 296.50 (9) & 3372.7 (1.0) & 0.18 & 5.3 & 1 & 10 &  & -0.070 &  \\
f14 & 297.44 (9) & 3362.0 (1.0) & 0.19 & 5.6 & 2 &  & 19 &  & 0.063 \\ \hline
\end{tabular}
\end{table*}

\begin{figure}
\centerline{\psfig{figure=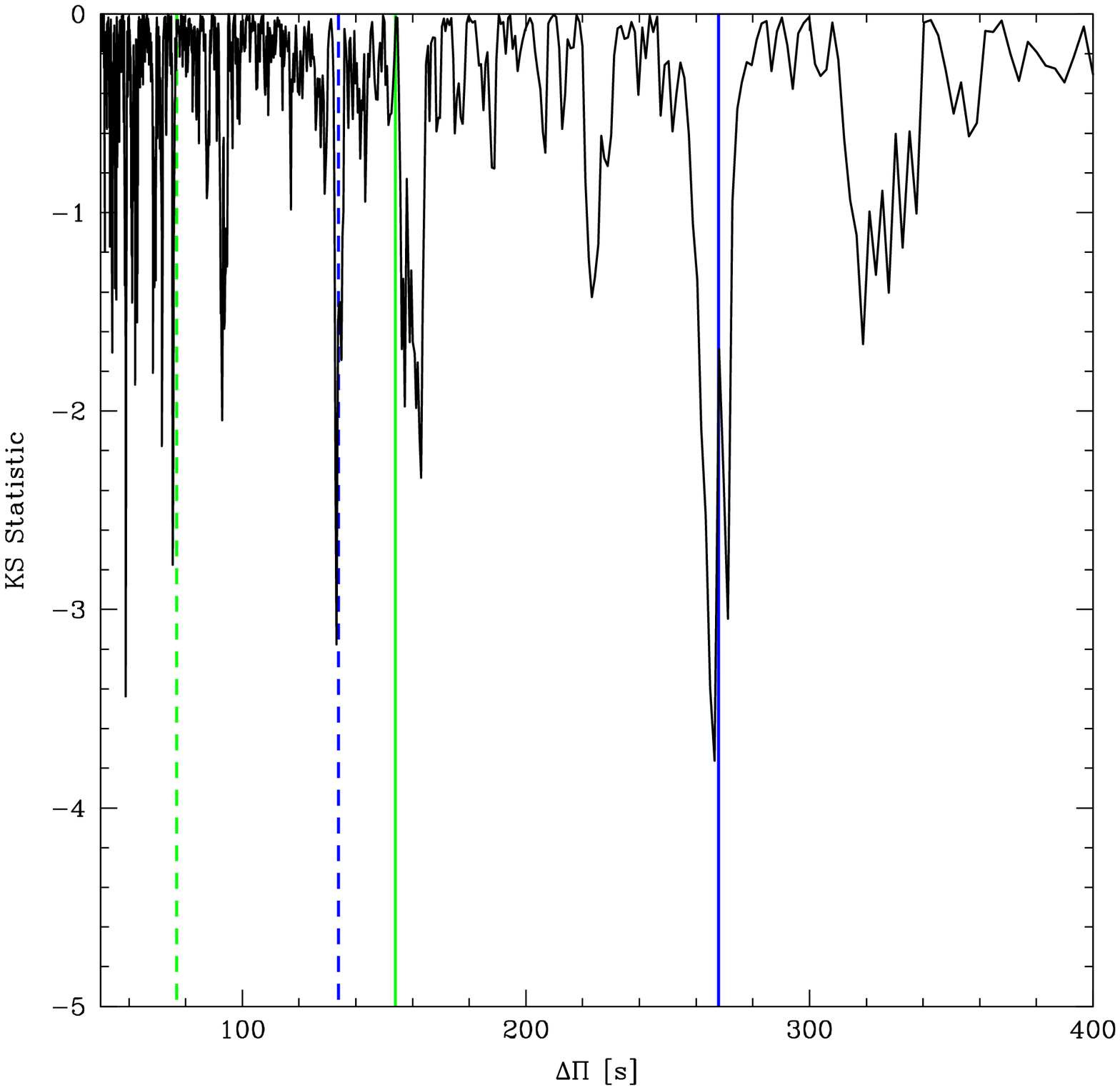,width=\columnwidth} }
\caption{Period spacing sequences indicated by the Kolmogorov-Smirnov test.
The solid blue (green) line indicates the $\ell\,=\,1$ (2)  period spacing sequence
with the dashed line the overtone alias.}
\label{fig02}
\end{figure}

\begin{figure*}
\centerline{\psfig{figure=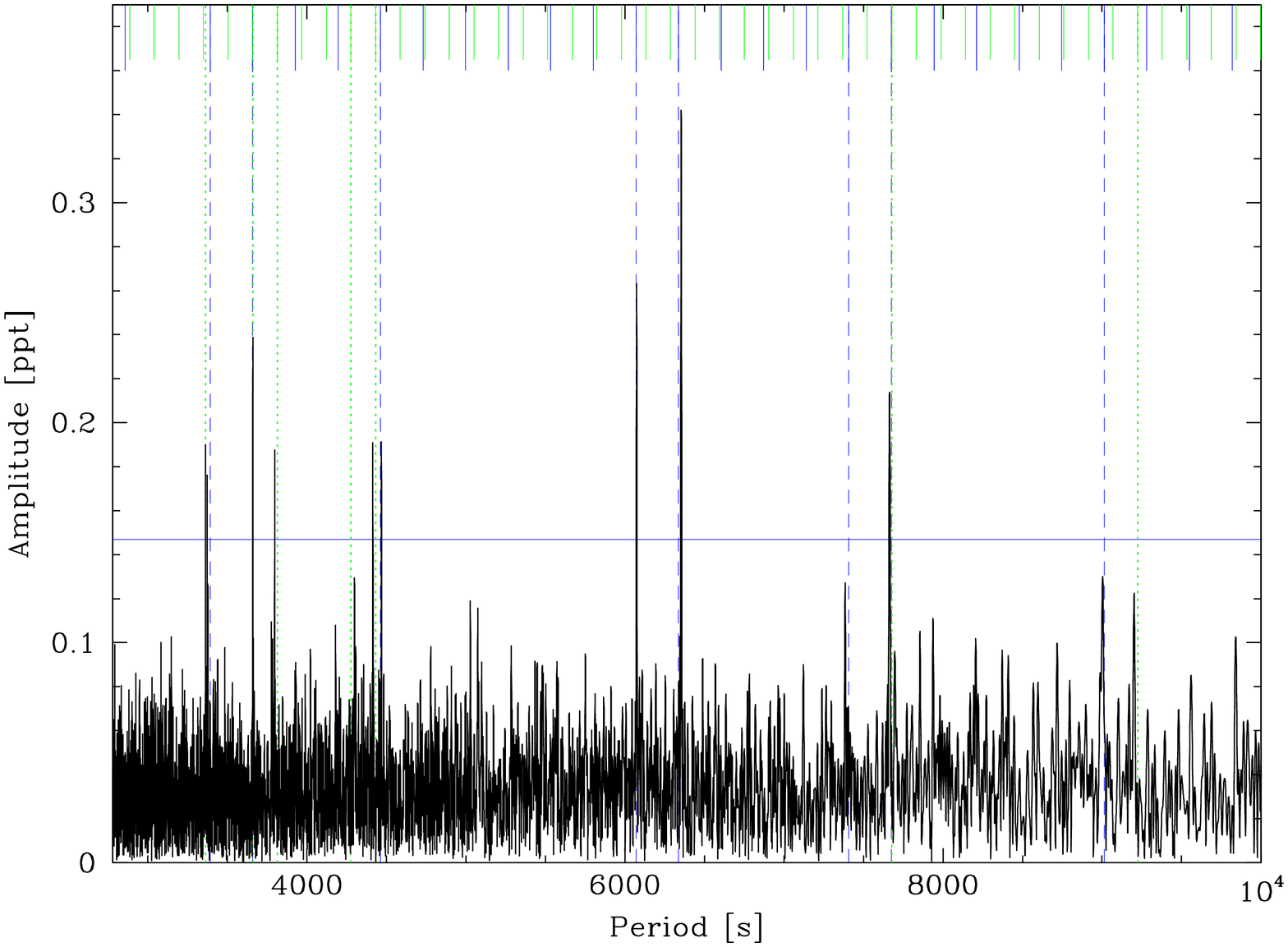,angle=-90,width=\textwidth} }
\caption{Period transform of PG\,1142 showing evenly spaced periods. Short
blue (green) lines indicate the asymptotic $\ell\,=\,1$ (2) sequence with
full-length lines indicating those which match detected peridicities. The
horizontal line indicates $4.35\sigma$ for reference.}
\label{fig03}
\end{figure*}

No consistent rotationally-induced frequency multiplets appear in these
data, so we can only use asymptotic period spacings for mode identifications.
A Kolmogorov-Smirnov (KS) test statistic was calculated using the periods
in Table\,\ref{tab01} with the results shown in Figure\,\ref{fig02}. From previous
work with sdBV stars, we know that the $\ell\,=\,1$ sequences should appear near 250\,s and the
$\ell\,=\,2$ sequence should appear near 150\,s \citep{reed11c}, 
and indeed the strongest KS statistic
appears near 265\,s. Differencing the periods indicated that most of PG\,1142's periods 
are separated nearly by this value or a multiple of it. The sequence was then determined
by assigning a relative overtone, $n$, and calculating a linear regression of those periods
and overtones. That fit was then used to find additional periods which fit the sequence. 
Nine of the 14 periods fit into this sequence with errors below 7\%.
According to asymptotic theory \citep[see ][]{seismobook},
the $\ell\,=\,2$ sequence
is related to the $\ell\,=\,1$ sequence as $\Delta\Pi_{\ell =2}\,=\,\Delta\Pi_{\ell =1}/
\sqrt{3}$ and so we could determine which periods should be $\ell\,=\,2$. Seven periods
could be associated with the $\ell\,=\,2$ sequence, including two peaks which also
fit the $\ell\,=\,1$ sequence. We cannot distinguish between $\ell\,=\,1$ or $2$ for
these peaks (f3 and f11)  and so Table\,\ref{tab01} has 
listings for both. Figure\,\ref{fig03} shows a period transform with the resultant
asymptotic period sequences indicated. The linear regression solutions
find period spacing sequences of $267.9\,\pm\,1.0$ and $153.9\,\pm\,0.7$\,seconds for 
the $\ell\,=\,1$ and $2$ sequences, respectively. This results 
in all periodicities being identifed with spacings
similar to other \emph{Kepler}-observed $g-$mode pulsators \citep{reed11c}.

\subsection{Binary Analysis}
\begin{figure*}
\centerline{\psfig{figure=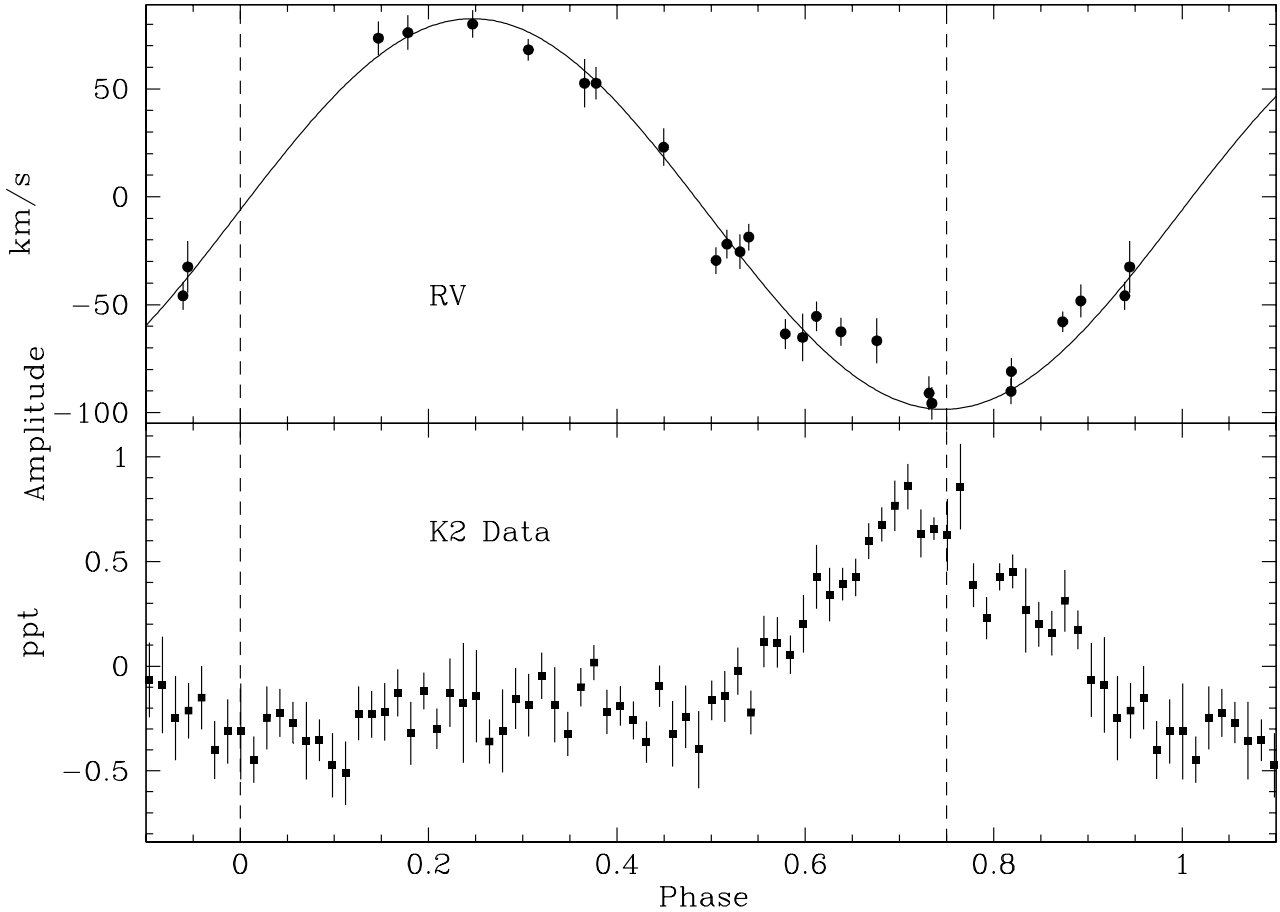,angle=-90,width=\textwidth} }
\caption{Velocity and photometric data folded over the orbital period. The K2
data have been phase-binned into 72 bins with $1\sigma$ error bars shown.}
\label{fig04}
\end{figure*}

PG\,1142 has proven to be a very exciting target. The detection of fA and fB prompted
follow-up spectroscopic observations (described in \S 2.1) to search for velocity
variations, which were subsequently detected. 
Fitting the 24 RVs (in Table\,\ref{tab02}), and assuming a circular orbit,
gives an orbital solution period of 
$0.541089\,\pm\,0.000020$\,d ($21.39\,\mu$Hz), amplitude of $85\,\pm\,2\,$km\,s$^{-1}$,
and phase of BJD$\,=\,2457099.74105\,\pm\,0.00204$ (top panel of
Fig.\,\ref{fig04}). The eccentricity was subsequently fitted at
$0.079\,\pm\,0.028$, showing it is negligible for our calculations. 
Using the values of $K\,=\,85\,\pm\,2\,$km\,s$^{-1}$ 
and P\,=\,0.541\,days from spectroscopy,
the sdB canonical mass M1=0.47 and assuming $i\,=\,90^o$,
 we get the minimum values for $M_2$ of $0.26\,M_{\odot}$ and for the orbital 
separation $a\,=\,a_1\,+\,a_2\,=\,2.52\,R_{\odot}$. Of course we know this is
not the inclination as no eclipses are evident in the lightcurve.

\begin{figure*}
\centerline{\psfig{figure=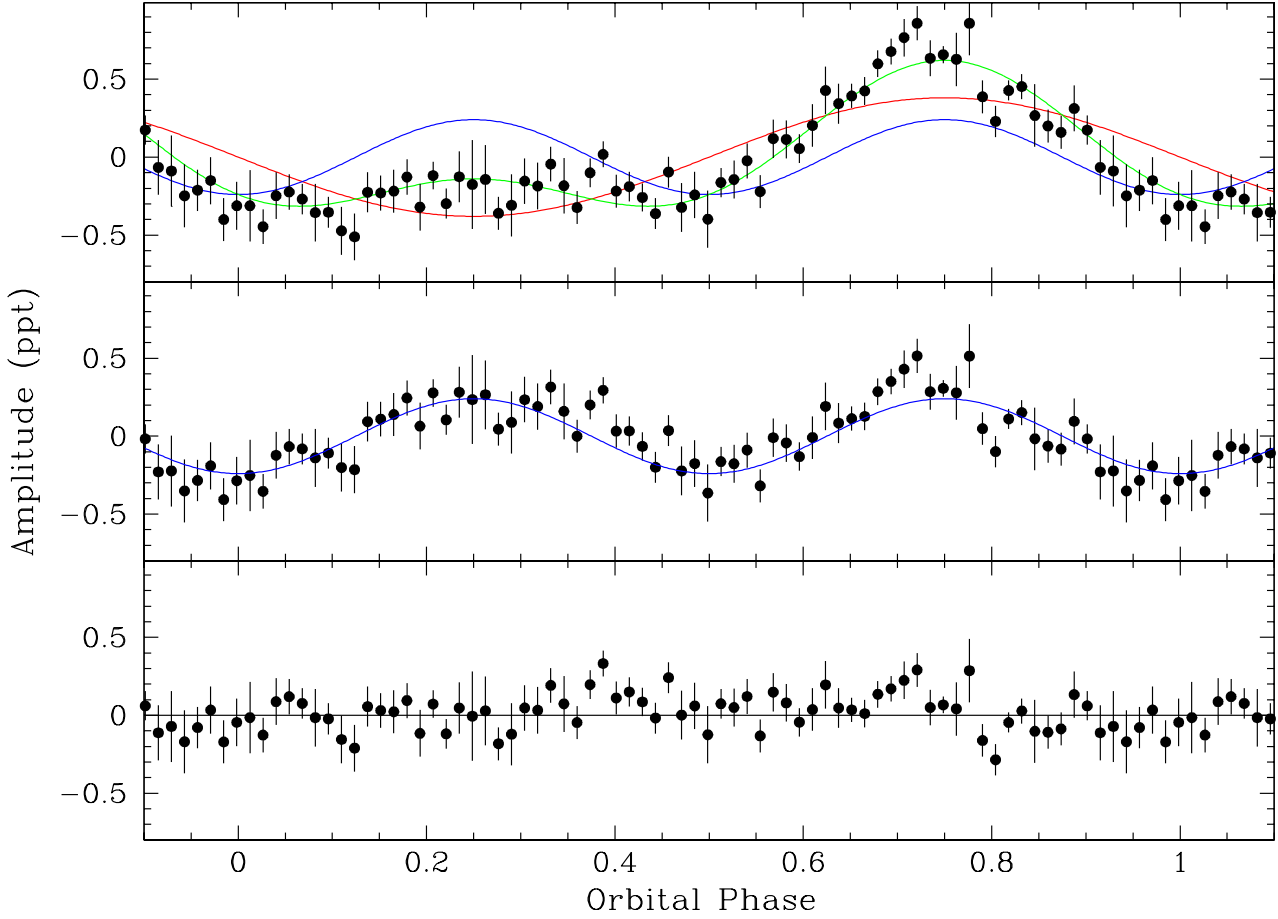,angle=-90,width=\textwidth} }
\caption{Orbital elements removed from the lightcurve. Top panel shows
original folded and binned lightcurve. Orbital fits are indicated by
lines with green being the combined fit,
blue the ellipsoidal deformation, and red the Doppler
boosting. Middle panel shows the Doppler-corrected 
lightcurve (with ellipsoidal deformation model) and the bottom panel
shows the residuals after both elements have been removed.}
\label{fig05b}
\end{figure*}

\begin{figure*}
\centerline{\psfig{figure=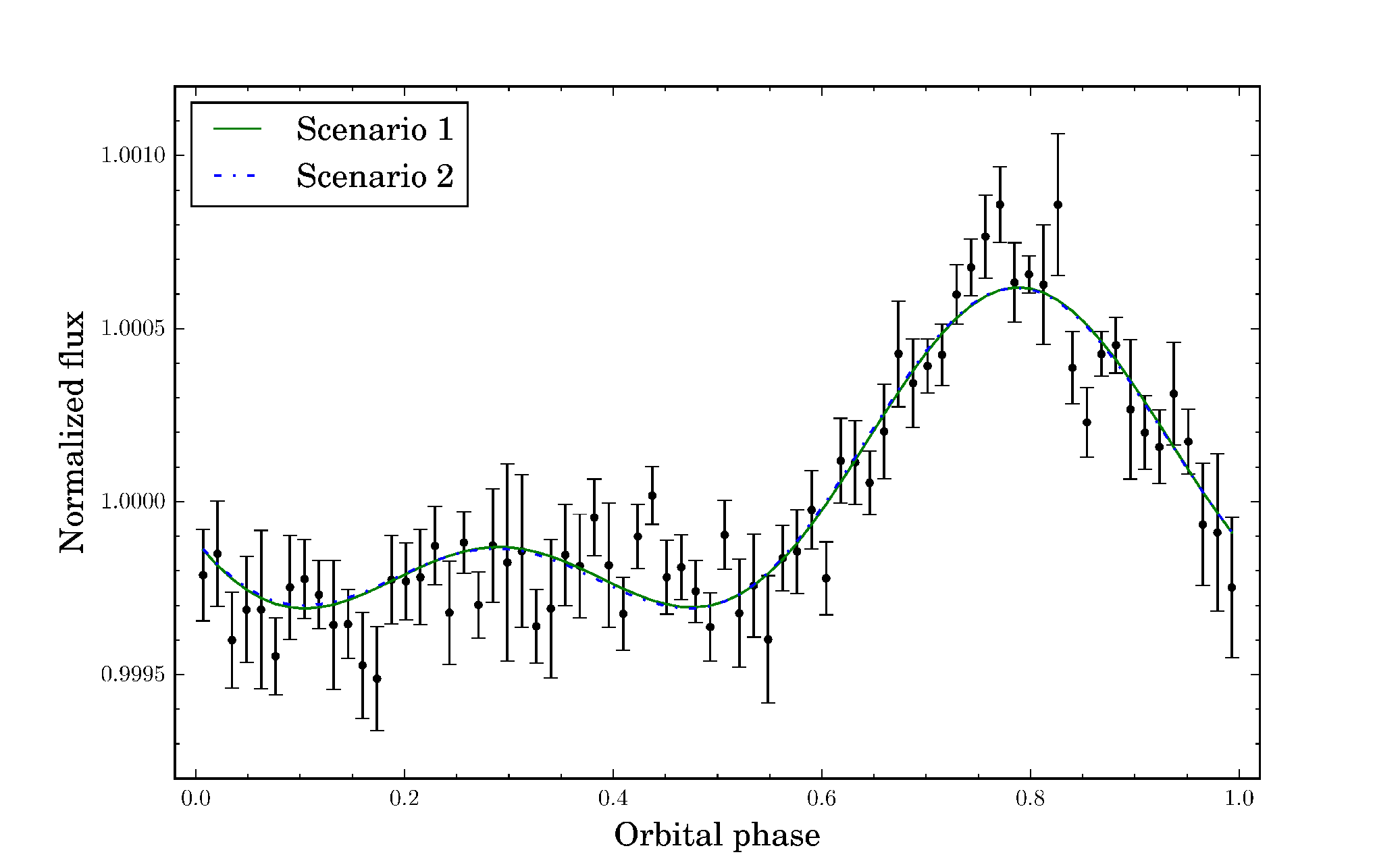,width=\textwidth} }
\caption{Model fits to the K2 lightcurve using the RV amplitude as
a constraint. As discussed in
\S 4, both fits use the canonical sdB mass of $0.47\,M_{\odot}$, Scenario 1 
has a white dwarf mass of $0.3\,M_{\odot}$ and Scenario 2 has a
white dwarf mass of $0.6\,M_{\odot}$. Both sets fit equally well.}
\label{fig06}
\end{figure*}

We attempted several phase-folding
and binning combinations of the photometry to improve the period
and phase, however the thruster firings, at roughly every six hours, and pulsations
decrease the accuracy, and so the RV data are better for constraining the period. 
Still, the photometry show
evidence of two-component variations, occuring
at similar levels, in agreement with the FT. From the phasing of the variations,
we can deduce the sources. A reflection-effect companion in the 
binary would produce a reflection effect
maximum, which would occur during a Doppler-boosting zero point (halfway between 
maximum and minimum) or an ellipsoidal-effect minimum.  A Doppler-boosting maximum 
would occur during an ellipsoidal-variable maximum. Figures\,\ref{fig05b} and \ref{fig06} 
show the
results of two-component fits to the folded lightcurve (described below). 
The phasing indicates that
the components are Doppler boosting and ellipsoidal variations. 
These two components indicate that PG\,1142 is most likely an sdB+WD binary. 

We do not notice any eclipses in
the folded lightcurve and the fitted Doppler and ellipsoidal amplitudes closely
match the peaks in the FT, indicating that there is no additional contribution
from eclipses. Using $R_{\rm sdB}\,=\,0.15\,R_{\odot}$ we calculate the inclination
to be less than $87^o$. As PG\,1142 is likely an sdB+WD binary, if we use the
canonical WD mass of $0.6\,M_{\odot}$, we calculate an inclination of $35^o$ and
$a\,=\,a_1\,+\,a_2\,=\,2.85\,R_{\odot}$

\begin{table}
\caption{Summary of spectroscopic and binary properties of PG\,1142.}
\label{tabbin}
\begin{tabular}{lll} \hline
Property & Value & Comments\\ \hline
$T_{\rm eff}$ & 27954(54)\,K & Combined spectrum\\
$\log g$ & 5.32(1)\,dex & Combined spectrum\\
$\log$(n(He)/n(H)) & -2.87(3)\,dex & Combined spectrum\\ \hline
Period & 0.541089(20)\,days & RVs, circular orbits\\
$K$ & 85(2)\,km\,s$^{-1}$ & RVs, circular orbits\\
Reduced $\chi^2$ of the fit & 1.4071 & RVs, circular orbits \\ 
Fit RMS & 7.8932 & RVs, circular orbits \\
$e$ & 0.079(28) & RVs \\ \hline
$\left(a_1\,+\,a_2\right)_{min}$ & $2.52R_{\odot}$ & $i=90^o$, $M_{\rm sdB}=0.47M_{\odot}$ \\
$\left(M_2\right)_{min}$ & $0.26M_{\odot}$ & $i=90^o$, $M_{\rm sdB}=0.47M_{\odot}$ \\ \hline
$i_{max}$ & $87^o$ & No observed eclipses.\\ \hline
$\left(a_1\,+\,a_2\right)_{canonical}$ & $2.85R_{\odot}$ & $M_{\rm sdB}=0.47M_{\odot}$, $M_{\rm WD}=0.60M_{\odot}$\\
$i_{canonical}$ & $35^o$ & $M_{\rm sdB}=0.47M_{\odot}$, $M_{\rm WD}=0.60M_{\odot}$ \\ \hline
$A_{ellipsoidal}$ & 0.25(2)\,ppt & Folded lightcurve fit.\\
$A_{Doppler}$ & 0.38(2)\,ppt & Folded lightcurve fit. \\
Reduced $\chi^2$ of the fit & 1.08 & Folded lightcurve fit. \\ \hline

\end{tabular}
\end{table}

A simple two-component sine fit was done to the phase-folded and binned lightcurve 
as shown in Fig.\,\ref{fig05b}. The resulting solution has a reduced $\chi^2=1.08$  
and fitted the ellipsoidal amplitude at $0.25\,\pm\,0.02\,$ppt
and the Doppler boosting at $0.38\,\pm\,0.02\,$ppt. From the RV amplitude 
of $85\,$km\,s$^{-1}$
and following the procedure of \citet{telting12a},
the Doppler boosting can be calculated to be 0.39\,ppt, which closely matches the 
photometric amplitude. The phasing of the Doppler and ellipsoidal
signals appear exactly as expected in phase: maximum Doppler boosting coincides
with an ellipsoidal maximum and the Doppler minimum with the other ellipsoidal
maximum. This indicates that we are correctly interpreting the causes of variations.

We have also modelled the binary light curve using the \texttt{LCURVE} 
code \citep{CopperwheatMarsh2010}. We tested two scenarios with very different component 
masses, but both consistent with the measured orbital period and radial velocity 
amplitude of the sdB. We assumed a white dwarf radius of $R_{\rm WD}= 0.013$ R$_\odot$ 
and an effective temperature of 27\,000 K for the sdB and 20\,000 K for the white 
dwarf. Limb darkening and gravity darkening coefficients do not have a large effect 
on the light curves and were fixed at the values used for KPD\,1946+4340 
in \citet{BloemenMarsh2011}. For Scenario 1, we used a white dwarf mass of 0.6 M$_\odot$ 
and an sdB mass of 0.47 M$_\odot$. These masses imply a mass ratio of $q=0.78$, and an 
inclination of $i=34.6$ deg. By fitting only the sdB radius, we found a best fit 
for $R_{\rm sdB}=0.228$ R$_\odot$. The best model has a reduced $\chi^2=1.13$, which 
indicates that the model nearly perfectly explains the features in the observed light 
curve. For Scenario 2, we changed the white dwarf mass to 0.3 M$_\odot$ and kept the 
sdB mass at the canonical value of 0.47 M$_\odot$, which implies a mass ratio of $q=1.57$ 
and a very different inclination of $i=65.6$ deg. We found a best fit 
for $R_{\rm sdB}=0.187$ R$_\odot$, resulting in a light curve with reduced $\chi^2=1.15$ 
which was nearly indistinguishable from the best fit in Scenario 1. Both fits are shown 
in Fig.\,\ref{fig06}. From this test, we conclude that sensible system parameters 
can explain the observed radial velocity curve and K2 light curve, but in the absence of 
eclipses, model fitting the light curve does not provide useful constraints on 
crucial parameters such 
as the white dwarf mass and the inclination of the system.

\section{Results and Discussion}
We discovered PG\,1142 to be a new sdB pulsator from 93\,days of K2 data
obtained during C1. In the photometry we
detect 14 periodicities associated with pulsations and two
others related to binarity. Using asymptotic period spacing, we were able to
identify all the pulsations as low-degree $\ell\,\leq\,2$ modes. 

\begin{figure*}
\centerline{\psfig{figure=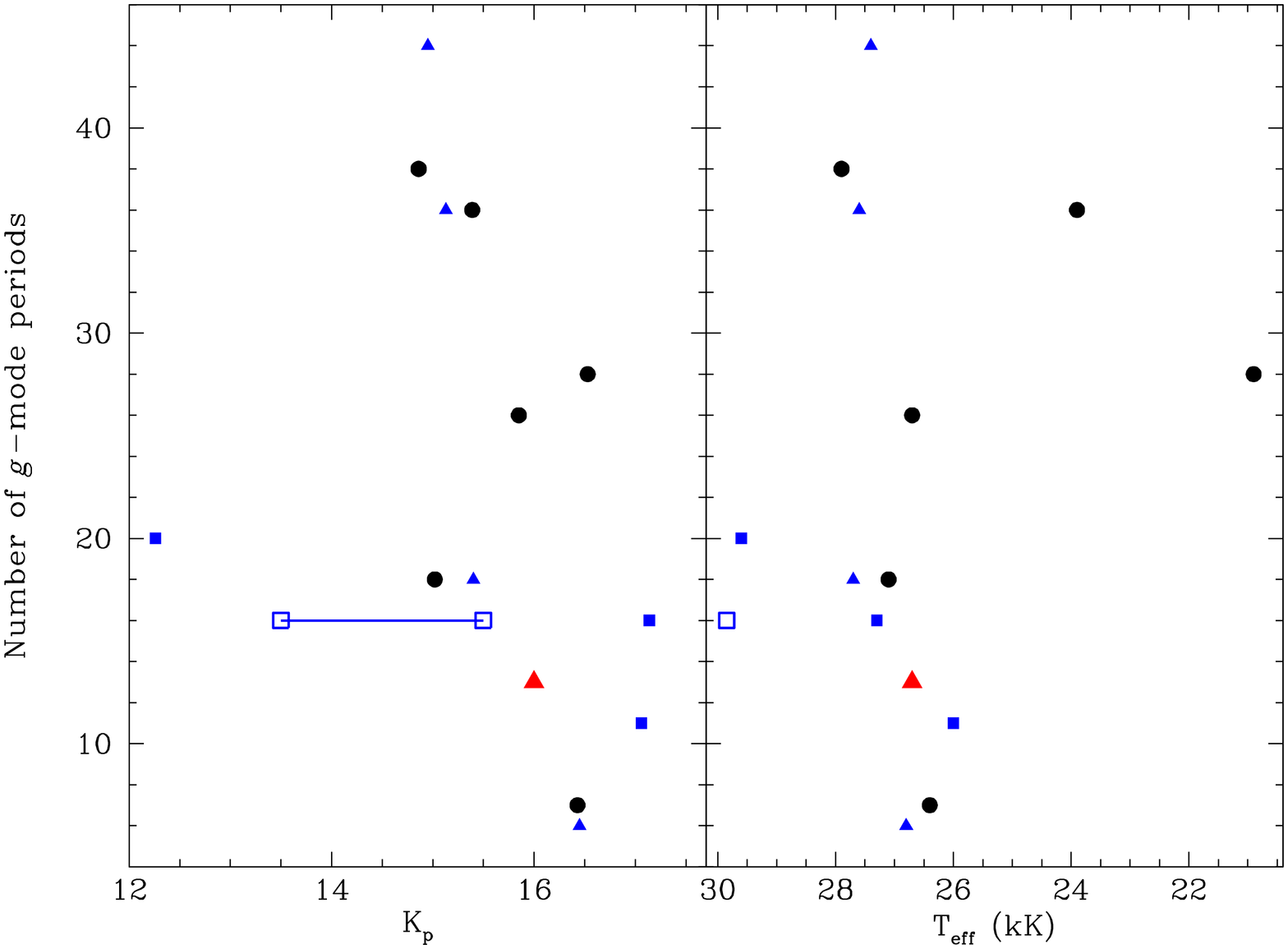,angle=-90,width=\textwidth} }
\caption{Number of $g-$mode pulsations detected with brightness and $T_{\rm eff}$
during \emph{Kepler's} survey phase \citep[summarized in][]{reed11c}, or K2's 
engineering campaign for EQ\,Psc \citep[open square,][]{csj14}. 
Squares indicate sdB+dM binaries, triangles 
sdB+WD binaries, and circles single stars. PG\,1142 is the red triangle. 
Note that the $K_p$ magnitude for EQ\,Psc is listed as 15.5
while it is known to be about 2 magnitudes brighter. Both values are indicated,
connected by a line.
}
\label{fig07}
\end{figure*}

One of our more surprising results is the \emph{lack} of periodicities in the data,
compared to other sdB pulsators.
With 93 nearly-continuous days of data, we were expecting to find more periodicities.
Of the 13 $g-$mode sdBV stars discovered using 30\,days of data during 
\emph{Kepler's} survey
phase \citep{roy10b,roy11b}, only three, all fainter than PG\,1142, 
had fewer periodicities and
even EQ\,Psc, using only nine days of K2 engineering data, shows more periodicities
\citep{csj14}.
Figure\,\ref{fig07} shows the number of $g-$mode pulsations detected from
\emph{Kepler} survey-phase and K2 engineering observations. 
It would be expected that fainter stars
would show fewer periods, having reduced S/N, and there is a very slight trend in
that direction. Flux from (non-white dwarf) companions could also reduce the number of pulsations
detected and the point types in the figure indicate those with companions. Again,
it is mildly suggestive that companion flux impacts the results. Position within
the instability strip could also affect the number of pulsations observed and so
the right panel indicates the number of pulsations observed against $T_{\rm eff}$. No
obvious trend occurs. 
While it was expected that K2 data would have more noise than the main
mission, these data contain three times as many points as survey-phase data, 
which should more than make
up for the reduced S/N of individual measurements. As such, we are left to 
conclude that PG\,1142 is one of the more sparse pulsators
observed by the \emph{Kepler} space telescope.

We also do not detect any obvious rotationally-induced frequency multiplets. If
PG\,1142 were tidally locked, the 0.54\,day orbit would produce $\ell\,=\,1$ and
2 splittings in multiples of 10.7 and 17.8\,$\mu$Hz, respectively. The frequency
differences appear roughly random and no frequency
differences are within 1\,$\mu$Hz of the expected separations. Since all the 
frequencies (with the possible exception of f12) fit period sequences, the most
likely explanation is that no frequency multiplets exist.
This is suggestive that PG\,1142 spins longer than $\sim45$\,days.
It is possible that f13 and the low-amplitude f12 form an $\ell\,=\,1$ doublet. If
this is the case, then the $0.40\,mu$Hz separation would indicate a spin period of
14.6\,days. While we feel this is unlikely and f12 is most likely a spurious frequency, 
it would still be subsynchronous to the binary period.
Of the nine
known rotation periods of sdBV stars observed with \emph{Kepler} and 
listed in Table\,1 of \citet[][and references therein]{me14},
four would surely have been detectable from K2 data with three more having
periods right around 45\,days, which likely would have been detected. Only two
have spin periods commensurate with the length of K2's C1, making it unlikely
multiplets would have been detectable with these data. From our experience with 
\emph{Kepler} and sdB variables, only extremely low-inclination angles suppress
multiplets \citep[see ][ for example]{reed04b} and for any reasonable range of
masses, the resultant orbital inclinations are well-suited for detecting rotationally-induced
multiplets. As such, the lack of observed multiplets indicate that
PG\,1142 has a long rotation period of at least 45\,days.

From spectroscopy and the folded lightcurve, we determined that PG\,1142 is in a
close binary with a period of 0.54\,d. We measured a Doppler boosting amplitude of 0.38\,ppt
which matches the orbital velocity of $85\,$km\,s$^{-1}$. We also measured
an ellipsoidal variable amplitude of 0.25\,ppt.
As we do not see eclipses,
the inclination must be $<87^o$ and, assuming canonical sdB and WD masses,
we calculated an inclination of $i\,=\,35^o$. At this inclination, mutiplets
should be easily observed if the rotation were commensurate with the
orbital period.
With the complete lack of observed multiplets implying a long rotation, this
system is an excellent test for various tidal locking mechanisms \citep[e.g.][]{pablo12}
and, so far as we know, it is the first ellipsoidal variable which is likely
rotating subsynchronously. With a rotation period $>45$\,days, 
PG\,1142 has the longest period of our known
subsynchronously rotating binaries; a particularly surprising result as it is
the only one with tidal deformation indicative of a stronger gravitational field
from the companion. 

Our results for PG\,1142
represent very useful data points to complete our understanding
of sdB stars, and by extension, the cores of red clump and other horizontal branch 
stars. The 14 observationally
identified modes will present challenges for model fitting. Even four years
after our first examination of \emph{Kepler} data, 
there are \emph{no} model fits using
observationally constrained mode identifications.
Once structural models accurately indicate
the internal structure (via asteroseismology matching), PG\,1142 and the other
subsynchronous binaries will be extremely useful for constraining mass and angular
momentum transfer.

ACKNOWLEDGMENTS: Funding for this research was provided by the National Science Foundation
grant\#1312869. Any opinions, findings, and conclusions or
recommendations expressed in this material are those of the
author(s) and do not necessarily reflect the views of the National
Science Foundation.
JK, LR, AW, and HF were supported by the Missouri 
Space Grant Consortium, funded by NASA.
ASB gratefully acknowledges a financial support from the Polish National Science Center under project No.\,UMO-2011/03/D/ST9/01914.
This paper includes data collected by the \emph{Kepler} mission. Funding for the
 \emph{Kepler} mission is provided by the NASA Science Mission directorate.
Data presented in this paper were obtained from the Mikulski Archive for 
Space Telescopes (MAST). STScI is operated by the Association of Universities 
for Research in Astronomy, Inc., under NASA contract NAS5-26555. Support 
for MAST for non-HST data is provided by the NASA Office of Space Science via 
grant NNX13AC07G and by other grants and contracts.

The spectroscopic observations used in this work were obtained with the
Nordic Optical Telescope at the Observatorio del Roque de los Muchachos
and operated jointly by Denmark, Finland, Iceland, Norway, and
Sweden and the
Mayall Telescope of Kitt Peak National Observatory, which is operated by the
Association of Universities for Research in Astronomy under cooperative
agreement with the National Science Foundation.

\bibliography{sdbrefsMNRAS}

\end{document}